\documentclass{PoS}

\title{Two-photon decay of $\pi^0$ from two-flavor lattice QCD}

\ShortTitle{Two-photon decay of $\pi^0$ from two-flavor lattice QCD}

\author{
  \speaker{E. Shintani}$^1$\thanks{shintani@riken.jp}, 
  S. Aoki$^2$, 
  S. Hashimoto$^{3,4}$
  T. Onogi$^5$ and 
  N. Yamada$^{3,4}$\\
  $^1$RIKEN-BNL Research Center, Brookhaven National Laboratory,
  Upton, NY 11973, USA\\ 
  $^2$Graduate School of Pure and Applied Sciences, University of
  Tsukuba, Tsukuba 305-8571, Japan\\ 
  $^3$KEK Theory Center, High Energy Accelerator Research Organization
  (KEK), Tsukuba 305-0801, Japan\\ 
  $^4$School of High Energy Accelerator Science, The Graduate
  University for Advanced Studies (Sokendai), Tsukuba 305-0801, Japan\\
  $^5$Department of Physics, Osaka University, Toyonaka 560-0043, Japan
        }
\abstract{
  We study the correction to the radiative $\pi^0$ decay width 
  due to finite light quark mass.
  Using lattice QCD with the overlap fermion formulation, we calculate
  the three-point function of the form $\langle PV_\mu V_\nu\rangle$
  in the (Euclidean) momentum space, which corresponds to the
  $\pi^0\rightarrow \gamma^*\gamma^*$ amplitude.
  To fit the lattice data, we use two different modifications of
  vector meson dominance (VMD) ansatz.
  One is a combined form of VMD with the next-to-leading order (NLO) 
  chiral perturbation theory (ChPT), and the other is a resummed form of
  pion-loop diagrams.  
  We extract one of the low energy constants in NLO ChPT, and 
  estimate $\pi^0\rightarrow \gamma\gamma$ decay width. 
}

\FullConference{The XXVIII International Symposium on Lattice Filed Theory\\
                 June 14-19,2010\\
                 Villasimius, Sardinia Italy}

\begin{document}

\section{Introduction}
We present an update on our study \cite{Shintani:2009qp} of the 
$\pi^0\rightarrow\gamma\gamma$ decay width using lattice QCD
calculation of its off-shell amplitude.
In the limit of massless up and down quarks, the corresponding
amplitude is completely determined by the triangle anomaly 
\cite{Adler:1969gk,Bell:1969ts},
which is valid to all orders in perturbation theory
\cite{Adler:1969er}.
Once the quark mass is introduced, the correction term arises, which
can be classified most conveniently in the effective field theory
framework \cite{Wess:1971yu,Witten:1983tw}.
The most recent analysis based on the next-to-next-to-leading order
chiral perturbation theory including the isospin breaking effect
\cite{Kampf:2009tk} quoted
$\Gamma_{\pi^0\gamma\gamma}$ = 8.09(11)~eV, which marginally agrees
with the recent experimental data
$\Gamma_{\pi^0\rightarrow\gamma\gamma}$ = 7.82(22)~eV
\cite{Larin:2010kq}.
The experimental error is expected to be reduced by a factor of 2 in
the near future, and the improvement of the theoretical prediction
would become more important.
Presently, the theoretical estimate is limited by the lack of
knowledge of the low energy constants in the effective theory, for
which the lattice input is most desirable.

We calculate the $\pi^0\rightarrow\gamma^*\gamma^*$ (two photons are
off-shell) amplitude in lattice QCD from the 
$\langle PV_\mu V_\nu\rangle$-type three-point function in the
momentum space. 
In our previous report \cite{Shintani:2009qp}, we showed that the 
$\pi^0\rightarrow\gamma^*\gamma^*$ amplitude calculated on the lattice
using the overlap fermion formulation is consistent with the
expectation from the axial anomaly, which is $1/(4\pi^2)$, after
extrapolating to the limit of on-shell photons and massless quarks.
The overlap fermion is most suitable for this study, because it has an
exact U(1)$_A$ chiral symmetry at the Lagrangian level that is violated by the
quantum effect.
Namely, it has the same symmetry structure as in the continuum theory
even at finite lattice spacings.

We work on a two-flavor QCD ensemble of a $16^3\times 32$ lattice with
the lattice cutoff $1/a$ = 1.667(17)~GeV \cite{Aoki:2008tq}.
Dynamical quarks are also described by the overlap fermion.
Topological charge $Q$ in this ensemble is fixed at $Q=0$, which 
is a source of the finite size effect \cite{Aoki:2007ka}.
There are four different quark masses, 
$m_q$ = 0.015, 0.025, 0.035, and 0.050 (in the lattice unit), 
which cover the range between
$m_s/6$ and $m_s/2$ with physical strange quark mass $m_s$. 

In the analysis presented last year \cite{Shintani:2009qp},
we used the naive vector meson dominance (VMD) ansatz to fit the
lattice data and to extrapolate to the limit of on-shell photons.
In the present report, we update this work including an analysis
using other functional forms motivated by the chiral effective theories.
We then present our preliminary result for the pion decay width.

Once the $\pi^0\rightarrow\gamma^*\gamma^*$ amplitude is determined
including its off-shell form factor, it may be used to estimate the
size of the light-by-light scattering contribution to the muon $g-2$,
which occurs through a (probably dominant) virtual process
$\gamma\gamma^*\to\pi^0\to\gamma^*\gamma^*$.


\section{$\pi^0\rightarrow \gamma^*\gamma^*$ transition form factor}
The $\pi^0\rightarrow\gamma^*\gamma^*$
transition form factor $f_{\pi^0\gamma^*\gamma^*}(p_1,p_2)$ is defined
as a matrix element of two electromagnetic (vector) currents $V_\mu^{\rm EM}$
between the pion state and the vacuum:
\begin{equation}
  \int\! d^4x_1 d^4x_2 e^{-ip_1x_1-ip_2x_2}
  \langle \pi^0(q)| V_\nu^{\rm EM}(p_2)V_\mu^{\rm EM}(p_1) |0\rangle = 
  \varepsilon_{\mu\nu\alpha\beta}p_1^\alpha p_2^{\beta}
  f_{\pi^0\gamma^*\gamma^*}(p_1,p_2),
  \label{eq:ff}
\end{equation}
where $p_1$ and $p_2$ denotes the off-shell photon momenta and $q=-p_1-p_2$.
On the lattice we extract $f_{\pi^0\gamma^*\gamma^*}(p_1,p_2)$ from a
three-point function
\begin{eqnarray}
  && G^{PVV}_{\mu\nu}(P_2,Q) = \sum_{x,y} e^{-iQx-iP_2y}
  \left\langle 2m_qP^{\rm rot\,3}(x)
    V^{\rm loc\,EM}_\nu(y)V^{\rm loc\,EM}_\mu(0)
  \right\rangle\nonumber\\ 
  &&
  =2m_q{\rm tr}[\tau^{3}Q_{e}^{2}]
  \left\langle\sum_{x,y,y'}e^{-iQx-iP_2y} 2{\rm Re}\,{\rm Tr}[
    S_{q}(0,x)\Gamma_P(x,y') S_{q}(y',y)\gamma_\nu
    S_{q}(y,0)\gamma_\mu]
  \right\rangle,  
  \label{eq:rhs}
\end{eqnarray}
where $\langle\cdots\rangle$ denotes an ensemble average.
$S_{q}(x,y)$ represents the quark propagator $D_{ov}(m)^{-1}(x,y)$
for the overlap-Dirac operator $D_{ov}(m)$ with the mass term $m$.
To achieve the $O(a)$-improvement of the pseudoscalar density operator
$P^{\rm rot\,3}$ contains the structure 
$\Gamma_{P}(x,y)=(1-D_{ov}(0)/M_0)\gamma_5(x,y)$.
``tr'' means a trace over flavor indices, and 
``Tr'' represents a trace over color and spinor indices.
The (Euclidean) four-momenta $P_1$, $P_2$ and $Q(=-P_1-P_2)$ are those
of two photons and a pion, respectively.
Their components are limited to discrete values 
$2\pi n_\mu/(L_\mu a)$, with the lattice spacing $a$ and 
the lattice size $L_{x,y,z}=16$ and $L_t=32$.
In obtaining the second line of (\ref{eq:rhs}), we assumed that the
disconnected quark diagrams are negligible.

With the overlap fermion formulation,
the (flavor non-singlet) pseudo-scalar density operator 
$P^{\rm rot\,3}(x)=\bar q(x)\tau^3\gamma_5[(1-D_{ov}(0)/M_0)q](x)$ 
satisfies the axial-Ward-Takahashi (AWT) identity 
$2m_qP^{\rm rot\,3}=\partial_\mu A_\mu^{\rm cv\,3}$
with the conserved axial-current $A_\mu^{\rm cv\,3}$
\cite{Kikukawa:1998py},
which we numerically confirmed.
For the electromagnetic current $V_\mu^{\rm loc\,EM}$ we use a local
current defined with the overlap fermion field. 
The electromagnetic charge is assigned as 
$Q_e={\rm diag}\{2/3e,-1/3e\}$ for up and down quarks.

Because we use the sequential source method, we have to invert the
overlap-Dirac operator for each choice of $P_2$, which is inserted at
the point $y$.
We restrict $P_2$ to the cases that it has a non-zero component only
in the temporal direction, $P_2=(0,0,0,P_{2t})$, with $\nu=1$.
Then we construct the three-point function by contracting the trace at
the point $x$ with a momentum $Q=-P_1-P_2$ inserted.
In order to minimize the numerical effort and the storage, we also
restrict $P_1$ as
\begin{eqnarray}
 P_{1} = (0,P_{1y},0,0)\textrm{ when $\mu=3$} & \textrm{  or  }& 
 P_{1} = (0,0,P_{1z},0)\textrm{ when $\mu=2$},
  \label{eq:P_2}
\end{eqnarray}
with all possible values of $P_{1y}$ and $P_{1z}$.
With these choices, we can pick up the non-zero components of 
$P_{1\alpha}P_{2\beta}\varepsilon_{\mu\nu\alpha\beta}$ 
in (\ref{eq:ff}).

We then extract the form factor $f_{\pi^0\gamma^*\gamma^*}(P_1,P_2)$ as
\begin{eqnarray}
  F^{\rm lat}(P_1,P_2) &=& G^{PVV}_{\mu\nu}(P_2,Q)\Big/\left(\sum_{\alpha\beta}
  P_{1\,\alpha}P_{2\,\beta}\varepsilon_{\mu\nu\alpha\beta}\right)
= -\frac{f_{\pi} m_\pi^2}{Q^2+m_\pi^2} f_{\pi^0\gamma^*\gamma^*}(P_1,P_2)
\label{eq:F^lat}
\end{eqnarray}
up to the contributions from excited states and disconnected diagrams.
Saturation of the pseudo-scalar channel by the pion field is quite
accurate for small momenta, because the first excited state, called
$\pi(1300)$ in PDG, is much heavier.
We numerically confirmed this by looking at the two-point function 
$\langle PP\rangle$ in the momentum space.

Since $P_1$ and $P_2$ are defined in the Euclidean space-time,
the on-shell condition is realized only at the zero momentum for both
$P_1$ and $P_2$.
In this limit, the three-point function trivially vanishes and no
information on the form factor could be extracted.
In order to obtain the decay width, we therefore have to work with
slightly off-shell photons and to extrapolate to the on-shell limit
assuming some functional form for $f_{\pi^0\gamma^*\gamma^*}(P_1,P_2)$.

\section{Fit ansatz}
For the form factor $f_{\pi^0\gamma^*\gamma^*}(P_1,P_2)$,
a naive choice would be that of the vector dominance model (VMD)
\begin{equation}
  \label{eq:VMD}
  F^{\rm VMD}(P_1,P_2)=-\frac{m_\pi^2}{Q^2+m_\pi^2}\frac{1}{4\pi^2}
  G_V(P_1,m_V)G_V(P_2,m_V),
\end{equation}
where $G_V(P,m_V)=m_V^2/(P^2+m_V^2)$ with the vector meson mass $m_V$.
In the limit of $P_1=P_2=0$, (\ref{eq:VMD}) reduces to $1/4\pi^2$,
which is the expectation of the anomaly valid in the massless limit.
Note that the above approximation is too naive as the same value $1/(4\pi^2)$ is also obtained 
even at finite quark mass, and so we clearly need modifications.

\begin{figure}[tb]
\begin{center}
  \includegraphics[width=30mm]{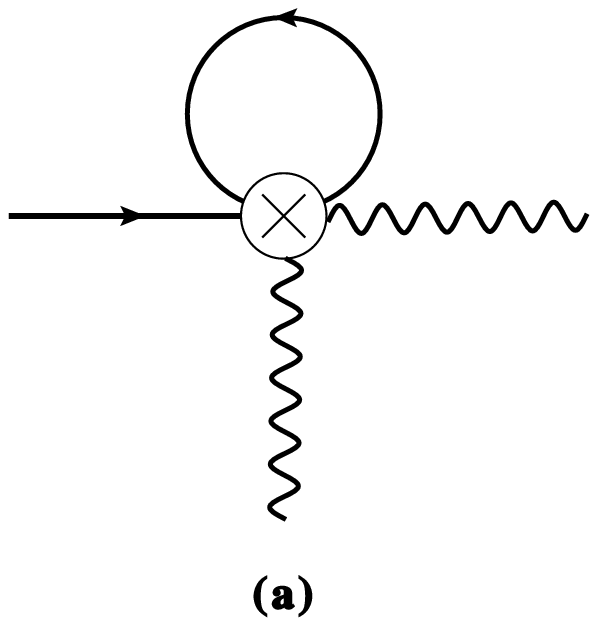}
  \hspace{3mm}
  \includegraphics[width=30mm]{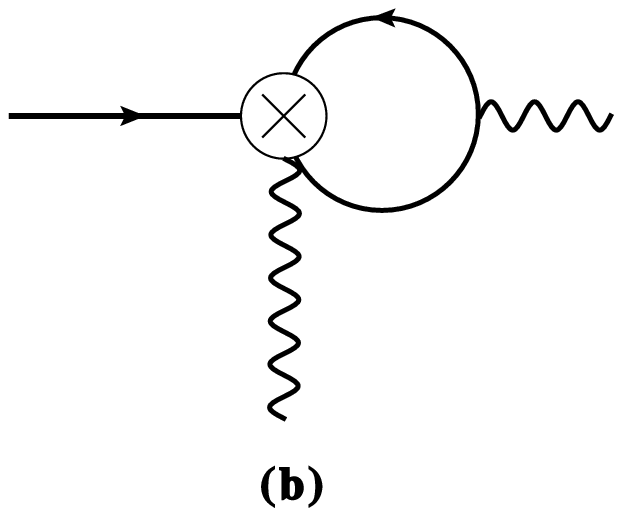}
  \hspace{3mm}
  \includegraphics[width=30mm]{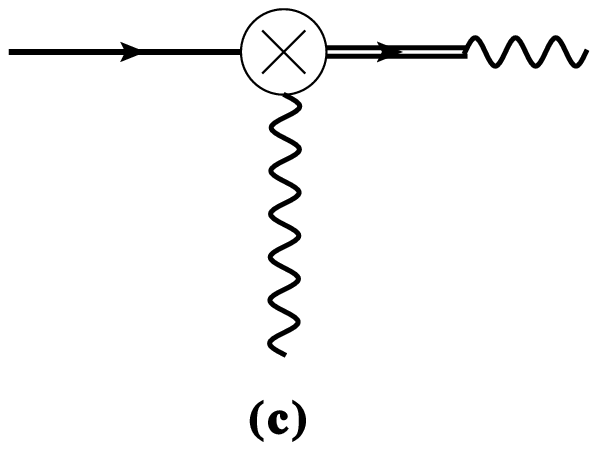}
  \hspace{3mm}
  \includegraphics[width=30mm]{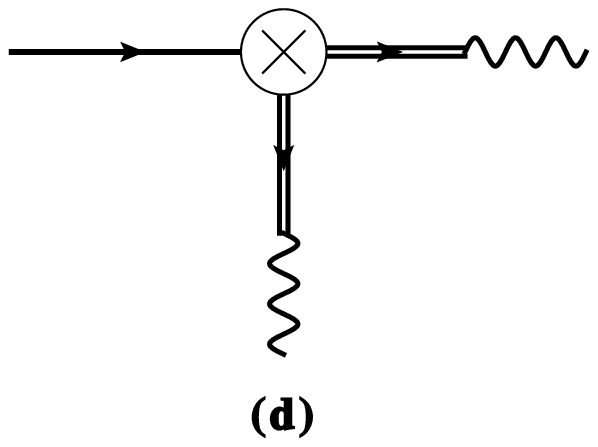}
  \caption{
    Vertices contribution to the $\pi^0\rightarrow\gamma^*\gamma^*$
    decay. 
    The solid, double-line and waved lines show pion, vector meson,
    and photon propagator, respectively. 
  } 
  \label{diagram1}
\end{center}
\end{figure}

Apart from this limit, we expect some corrections.
For small quark mass and external momenta, those can be described by
the one-loop ChPT.
A formula for the cases of off-shell photons is available in 
\cite{Borasoy:2003yb}.
We slightly extend the formula by adding some analytic term and use
a function
\begin{eqnarray}
  \Gamma^{\rm 1loop+VMD}(m_\pi^2,P_1^2.P_2^2) &=&  1 + \frac{1}{f_\pi^2}\Big(
  -\frac{4}{3}\Delta_\pi(m_\pi^2) + J(m_\pi^2,P_1^2) + J(m_\pi^2,P_2^2)\Big) 
  \nonumber\\
  & & - \frac{256\pi^2}{3}m_\pi^2 y_1 
  + \frac{64\pi^2}{3}\left(\frac{P_1^2}{P_1^2+m_V^2}+\frac{P_2^2}{P_2^2+m_V^2}\right)y_2 \nonumber\\
  & & + \frac{P_1^2P_2^2}{(P_1^2+m_V^2)(P_2^2+m_V^2)}y_3
  \label{eq:F^VMDChPT}
\end{eqnarray}
for $-4\pi^2 F(P_1,P_2)/(m_\pi^2/(Q^2+m_\pi^2))$.
(Namely, $\Gamma(m_\pi^2,P_1^2,P_2^2)$ becomes 1 in the massless and
soft photon limit.)
Here, the first line in the above equation corresponds to the one-loop
corrections.
The tadpole diagram (Figure~\ref{diagram1}(a)) and the vector-vector
correlator diagram (Figure~\ref{diagram1}(b)) lead to the following
non-analytic functions 
\begin{eqnarray}
  \Delta_\pi(m_\pi^2) &=& \frac{m_\pi^2}{16\pi^2}\ln \frac{m_\pi^2}{\mu^2},\\
  J(m_\pi^2,P^2) &=& 
  \frac{2}{3}\left[ \frac{P^2}{64\pi^2}\sigma^2\left\{
      -1 + \ln\frac{m_\pi^2}{\mu^2} +
      2\sigma\tanh^{-1}\frac{1}{\sigma}
    \right\} - \frac{P^2+6m_\pi^2}{96\pi^2} 
  \right],
\end{eqnarray}
respectively, with $\sigma=\sqrt{1+4m_\pi^2/P^2}$.
The scale $\mu$ is the renormalization point that defines the counter terms. 
We set $\mu$ = 775~MeV, the physical $\rho$ meson mass.

The second and third lines of (\ref{eq:F^VMDChPT}) define the analytic (or counter) terms.
Since the lowest non-zero momentum $P_1^2\simeq$ 0.43~GeV$^2$ is
already quite large to apply the one-loop ChPT formula, we modify the
analytic terms motivated by VMD.
Namely, the term with a single vector meson propagator (the forth term) represents the effective
interaction of the form $\pi^0\gamma\rho$ (Figure~\ref{diagram1}(c)),
while the last term represents $\pi^0\rho\omega$ (Figure~\ref{diagram1}(d)).
The parameters $y_1$, $y_2$ and $y_3$ are to be determined by a
numerical fit of the lattice data.

\begin{figure}
\begin{center}
  \includegraphics[width=60mm]{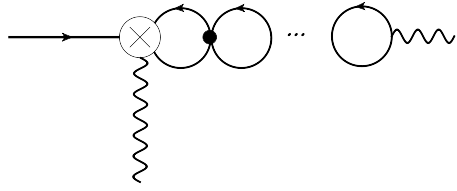}
  \caption{
    Bubble diagram of pion loops. 
    Symbols are the same as in Figure~\protect\ref{diagram1}. 
    Filled circle shows the tree-level vertex for the $\pi$-$\pi$
    scattering diagram.
  } 
  \label{diagram2}
\end{center}
\end{figure}

Another fit form we attempted is that of a resummed pion loops.
Again motivated by VMD, we assume that the bubble diagrams shown in
Figure~\ref{diagram2} dominate the higher loop corrections and
eventually produce the vector meson pole after summed up.
Then, an expected functional form in the channel corresponding to the
rho meson would be
\begin{eqnarray}
  \Gamma^{\rm resum}(m_\pi^2,P_1^2.P_2^2) &=& 1-\frac{2}{f_\pi^2}\Delta_\pi(m_\pi^2)
  -\frac{256\pi^2}{3}m_\pi^2 z_1\nonumber\\
 &+& \frac{L(m_\pi^2,P_1^2,z_2)}{1-2 L(m_\pi^2,P_1^2,z_2)}
  +  \frac{L(m_\pi^2,P_2^2,z_2)}{1-2 L(m_\pi^2,P_2^2,z_2)},
  \label{eq:F^resum}\\
  L(m_\pi^2,P^2,z_2) &=& \frac{1}{f_\pi^2}
  \left( \frac{1}{3}\Delta_\pi(m_\pi^2) + J(m_\pi^2,P^2) \right) 
  + \frac{64\pi^2}{3}P^2z_2,
\end{eqnarray}
where $z_1$ and $z_2$ are fit parameters.

\section{Fitting results}
Here we describe the fit of the lattice data.

\begin{figure}[tb]
\begin{center}
  \vskip 5mm
  \includegraphics[width=85mm]{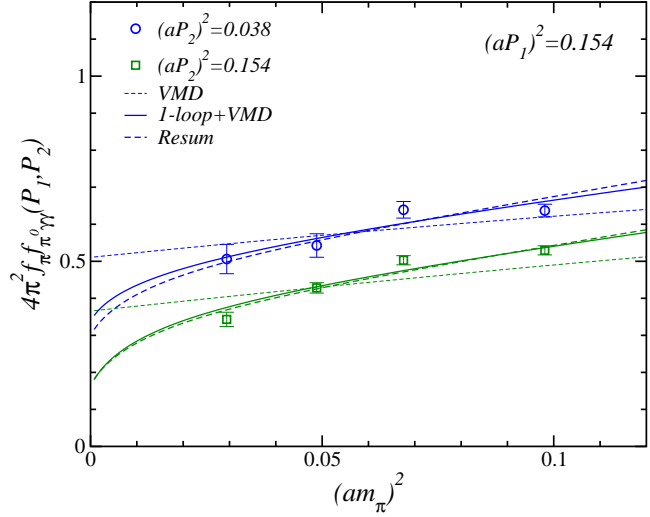}
  \caption{
    $\pi^0\rightarrow\gamma^*\gamma^*$ transition form factor as a
    function of pion mass squared.
    Different symbol denotes the different $(aP_2)^2$, and 
  $(aP_1)^2$ is fixed in the lowest value, $0.154$.
  Thick solid and dashed curves are results in fit function $\Gamma^{\rm 1loop+VMD}$
  and $\Gamma^{\rm resum}$ respectively. 
  Thin dashed curve shows the naive VMD results as the comparison with them.
  }
  \label{fig:G^lat}
\end{center}
\end{figure}

In order to keep the external momenta as small as possible, we
restrict the momentum range to use, so that only 
$(aP_1)^2=0.154$ and $0.039\le(aP_2)^2\le 0.154$ are included. 
These correspond to the lowest $(aP_1)^2$ and the lowest and second
lowest $(aP_2)^2$ in the momentum assignment (\ref{eq:P_2}). 
The data points are plotted in Figure~\ref{fig:G^lat} as a function of pion mass squared.
The plot is normalized so that the on-shell limit corresponds to 1 in the chiral limit.
So, we can see that the deviation from that limit is quite significant
(as large as $\sim$ 50\%).

By fitting the lattice data with the functions 
$\Gamma^{\rm 1loop+VMD}(m_\pi^2,P_1^2,P_2^2)$ and 
$\Gamma^{\rm resum}(m_\pi^2,P_1^2,P_2^2)$,  
we obtain the solid and dashed curves shown in
Figure~\ref{fig:G^lat}, respectively. 
Here we input the measured value of $m_V$ 
from two-point function $\langle VV\rangle$ 
into $\Gamma^{\rm 1loop+VMD}(m_\pi^2,P_1^2,P_2^2)$.
The naive VMD is also plotted by thin dashed curves for comparison.
In the plot, one can see that both fit curves with the modified VMD
show a better agreement with the lattice data than the naive VMD does,
especially for the second lowest momentum.

\begin{table}[tb]
\begin{center}
\begin{tabular}{cccc}
\hline\hline
 & $y_1(=w_3)$ & $y_2$ & $y_3$ \\
\hline
1-loop + VMD & $-$0.00030(20) GeV$^{-2}$ & $-$0.0054(6) & 1.65(48) \\
\hline\hline
 & $z_1(=w_3)$ & $z_2$  \\
\hline
Resummation & $-$0.00120(12) GeV$^{-2}$ & $-$0.0271(37) GeV$^{-2}$ & \\
\hline\hline
\end{tabular}
\caption{Fitting results with the function 
  $\Gamma^{\rm 1loop+VMD}(m_\pi^2,P_1^2,P_2^2)$ and with
  $\Gamma^{\rm resum}(m_\pi^2,P_1^2,P_2^2)$. 
  $y_1(z_1)$ corresponds to LECs, $w_3$. 
  }
\label{table1}
\end{center}
\end{table}

Fit parameters are listed in Table~\ref{table1}.
A comparison with the phenomenological analysis in
\cite{Borasoy:2003yb} is possible for the parameter controlling the
pion mass dependence $w_3$, which corresponds to $y_1$ and $z_1$ in
our functional form.
Unfortunately, we find a significant discrepancy between $y_1$ and $z_1$.
This suggests that the extraction of the low energy constants in the
chiral effective theory from our calculation suffers from large
systematic error.
That is due to several possible reasons. 
One of them appears to be due to the large momentum
we took for the off-shell photons to apply the one-loop ChPT.
we estimate this parameter as
$w_3 = -0.00075(23)(63) \,{\rm GeV}^{-2}$,
where the first error is statistical the second error denotes 
the systematic error coming from the ambiguity of the fit function.
The phenomenological work used a value $w_3=0.0131$
\cite{Borasoy:2003yb}, which is an order of magnitude larger than our result. 

The $\pi^0\rightarrow \gamma\gamma$ decay width is defined as 
\begin{equation}
  \Gamma_{\pi^0\gamma\gamma} = \frac{m_\pi^3\alpha_e^2}{64\pi^3f_\pi^2}
  |\Gamma^{\rm fit}_{\textrm{on-shell}}|^2
\end{equation}
with QED fine structure constant $\alpha_e=1/137$.
Inserting the fitting results
after taking $P_1^2=P_2^2=0$, which 
is consistent with CHPT formula in on-shell two photon decay, 
we obtain
\begin{equation}
  \Gamma_{\pi^0\gamma\gamma} = 7.98(6)(20)\,{\rm eV},
\end{equation}
where the first error is statistical and the second is systematic one.
While the systematic uncertainty is still rather large in our
calculation, the result and its error are compatible with the present
world average, 7.87(47)~eV \cite{Nakamura:2010zzi},  
and also with the recent experimental result 7.82(22)~eV
\cite{Larin:2010kq}.

\section{Conclusions}
Lattice calculation of the $\pi^0\rightarrow\gamma^*\gamma^*$ decay
amplitude is feasible for space-like momenta $p_1$, $p_2$ and $q$.
We attempt a fit of the lattice data with the functional forms
motivated by the NLO ChPT, from which we can in principle extract the
low energy constants in the effective theory.
With our current setup, however, the smallest non-zero momentum is
already too large to safely apply the NLO ChPT, so that we have to
introduce some model functions to extend the ChPT form to larger momenta.
Our proposals partly motivated by the vector meson dominance ansatz
describe the lattice data very well, but are not an unique choice.
The result for the physical decay width thus has large systematic
effect, but is already compatible with other estimates in the size of uncertainty.
An extension to larger lattices and an introduction of the twisted
boundary condition will improve this situation a lot, which is a
subject of future studies.

Numerical calculations are performed on IBM System Blue Gene Solution
and Hitachi SR11000 at High Energy Accelerator Research Organization (KEK) 
under a support of its Large Scale Simulation Program (No.~09/10-09).
This work is supported by the Grant-in-Aid of the Japanese Ministry of
Education 
(No. 20105002, 
     21105508, 
     21764002,
     22011012, 
     22740183  
).


\begin{thebibliography}{99}

\bibitem{Shintani:2009qp}
  E.~Shintani, S.~Aoki, S.~Hashimoto, T.~Onogi and N.~Yamada  [JLQCD
                  Collaboration],
  PoS {\bf LAT2009}, 246 (2009)
  [arXiv:0912.0253 [hep-lat]].

\bibitem{Adler:1969gk}
  S.~L.~Adler,
  Phys.\ Rev.\  {\bf 177}, 2426 (1969).

\bibitem{Bell:1969ts}
  J.~S.~Bell and R.~Jackiw,
  Nuovo Cim.\  A {\bf 60}, 47 (1969).

\bibitem{Adler:1969er}
  S.~L.~Adler and W.~A.~Bardeen,
  Phys.\ Rev.\  {\bf 182}, 1517 (1969).

\bibitem{Wess:1971yu}
  J.~Wess and B.~Zumino,
  Phys.\ Lett.\  B {\bf 37}, 95 (1971).

\bibitem{Witten:1983tw}
  E.~Witten,
  Nucl.\ Phys.\  B {\bf 223}, 422 (1983).

\bibitem{Kampf:2009tk}
  K.~Kampf and B.~Moussallam,
  Phys.\ Rev.\  D {\bf 79}, 076005 (2009)
  [arXiv:0901.4688 [hep-ph]].

\bibitem{Larin:2010kq}
  I.~Larin {\it et al.}  [PrimEx Collaboration],
  arXiv:1009.1681 [nucl-ex].

\bibitem{Nakamura:2010zzi}
  K.~Nakamura {\it et al.}  [Particle Data Group],
  J.\ Phys.\ G {\bf 37}, 075021 (2010).









\bibitem{Aoki:2008tq}
  S.~Aoki {\it et al.}  [JLQCD Collaboration],
  Phys.\ Rev.\  D {\bf 78}, 014508 (2008)
  [arXiv:0803.3197 [hep-lat]].


\bibitem{Aoki:2007ka}
  S.~Aoki, H.~Fukaya, S.~Hashimoto and T.~Onogi,
  Phys.\ Rev.\  D {\bf 76}, 054508 (2007)
  [arXiv:0707.0396 [hep-lat]].

\bibitem{Kikukawa:1998py}
  Y.~Kikukawa and A.~Yamada,
  Nucl.\ Phys.\  B {\bf 547}, 413 (1999)
  [arXiv:hep-lat/9808026].

\bibitem{Borasoy:2003yb}
  B.~Borasoy and R.~Nissler,
  Eur.\ Phys.\ J.\  A {\bf 19}, 367 (2004)
  [arXiv:hep-ph/0309011].










\end{thebibliography}
\end{document}